
\hoffset=0.7cm \voffset=0.2cm
\vbadness=10000
%
%

\font\sss=cmssq8 scaled 1000
\font\bf=cmbx10 scaled 1200
\font\bb=cmbx10 scaled 1920

\font\rs=cmr8 scaled 1200
\font\it=cmti10 scaled 1200

\font\sc=cmcsc10 scaled 1200
\font\tenrm=cmr10 scaled 1200
\font\sevenrm=cmr7 scaled 1200
\font\fiverm=cmr5 scaled 1200
\font\teni=cmmi10 scaled 1200
\font\seveni=cmmi7 scaled 1200
\font\fivei=cmmi5 scaled 1200
\font\tensy=cmsy10 scaled 1200
\font\sevensy=cmsy7 scaled 1200
\font\fivesy=cmsy5 scaled 1200

\font\tenbf=cmbx10 scaled 1200
\font\sevenbf=cmbx7 scaled 1200
\font\fivebf=cmbx5 scaled 1200
\font\tensl=cmsl10 scaled 1200
\font\tentt=cmtt10 scaled 1200
\font\tenit=cmti10 scaled 1200
\catcode`\@=11
\textfont0=\tenrm \scriptfont0=\sevenrm \scriptscriptfont0=\fiverm
\def\rm{\fam\z@\tenrm}
\textfont1=\teni \scriptfont1=\seveni \scriptscriptfont1=\fivei
\def\mit{\fam\@ne} \def\oldstyle{\fam\@ne\teni}
\textfont2=\tensy \scriptfont2=\sevensy \scriptscriptfont2=\fivesy
\def\cal{\fam\tw@}
\textfont3=\tenex \scriptfont3=\tenex \scriptscriptfont3=\tenex
\newfam\itfam \def\it{\fam\itfam\tenit} 
\textfont\itfam=\tenit
\newfam\slfam  
\textfont\slfam=\tensl
\newfam\bffam \def\bf{\fam\bffam\tenbf} 
\textfont\bffam=\tenbf \scriptfont\bffam=\sevenbf
\scriptscriptfont\bffam=\fivebf
\newfam\ttfam  
\textfont\ttfam=\tentt
\catcode`\@=12
%
%
\rm


\abovedisplayskip=30pt plus 4pt minus 10pt
\abovedisplayshortskip=20pt plus 4pt
\belowdisplayskip=30pt plus 4pt minus 10pt
\belowdisplayshortskip=28pt plus 4pt minus 4pt
\def\folio{\ifnum\pageno=1\nopagenumbers\else\number\pageno\fi}

\def\--{\! - \!}

\hfuzz=10pt \overfullrule=0pt
\vsize 8.75in
\hsize 6in
\baselineskip=14pt

\parindent 20pt \parskip 6pt

\def\blankline{\par\vskip \baselineskip}
\def\twocol#1{\halign{##\quad\hfil &##\hfil\cr #1}}

 \mathcode`*="002A

\def\_{\vrule height 0.8pt depth 0pt width 1em}

\newbox\grsign \setbox\grsign=\hbox{$>$} \newdimen\grdimen \grdimen=\ht\grsign
\newbox\simlessbox \newbox\simgreatbox
\setbox\simgreatbox=\hbox{\raise.5ex\hbox{$>$}\llap
     {\lower.5ex\hbox{$\sim$}}}\ht1=\grdimen\dp1=0pt
\setbox\simlessbox=\hbox{\raise.5ex\hbox{$<$}\llap
     {\lower.5ex\hbox{$\sim$}}}\ht2=\grdimen\dp2=0pt

\def\simless{\mathrel{\copy\simlessbox}}

\def\cfalh{\par\vfil\eject \vskip -12pt \moveleft 0.5in\vbox{
     \twocol{{\bb Center for Astrophysics}\hbox to 1.5in{} &\cr
     {\sss 60 Garden Street} & {\sss Harvard College Observatory} \cr
     {\sss Cambridge, Massachusetts 02138} & {\sss Smithsonian
          Astrophysical Observatory}\cr}}\par\blankline}

\def\listitem{\par \hangindent=50pt\hangafter=1
     $\ $\hbox to 20pt{\hfil $\bullet$ \hfil}}
\def\date#1{\par\hbox to \hsize{\hfil #1\qquad}\par}

\def\refitem{\par\parskip 0pt\noindent\hangindent 20pt}
\def\ref#1{$^{#1}$}
\def\title#1\endtitle{\par\vfil\eject
     \par\vbox to 1.5in {}{\bf #1}\par\vskip 1.5in\nobreak}
\def\author#1\endauthor{\par{\sc #1}\par\blankline}
\def\institution#1\endinstitution{{\rs #1}}

\def\section#1\endsection{\par\vfil\eject{\bf #1}\par\vskip 12pt\nobreak\rm}
\def\subsection#1\endsubsection{\vskip 14pt plus 50pt {\rm #1}\par
     \nobreak\blankline\nobreak\rm}

\phantom{.}
\vskip 1.0truein
\centerline{\bf X-RAY AND OPTICAL PROPERTIES OF}
\centerline{{\bf GROUPS OF GALAXIES}$^1$}
\vskip 1.0truein
\centerline{Ian P. Dell'Antonio, Margaret J. Geller, Daniel G. Fabricant}
\centerline{Harvard-Smithsonian Center for Astrophysics,60 Garden St., }
\centerline{Cambridge, Ma, 02138.}
\vskip 3.0truein
$~^1$ Observations reported here were obtained at the F. L. Whipple Observatory
and at the Multiple Mirror Telescope, a joint facility of the Smithsonian
Institution and the University of Arizona.
\par\vfil\eject
\centerline{\bf ABSTRACT}
\vskip 0.3 in
We have measured $125$ redshifts in 31 groups of galaxies observed with
Einstein, and
have compiled an additional 543 redshifts from the literature.
There is a correlation betweeen galaxy surface density and
 group velocity dispersion, with $\mu \propto \sigma^{1.6\pm 0.6}$, but
the scatter about this relation is large.

We examine the relationship between the
group x-ray luminosity in the 0.3-3.5 keV band and the measured velocity
dispersion.  Richer
groups follow the same relation as rich clusters (cf. Quintana \& Melnick 1982)
with $L_x \propto \sigma^{4.0\pm 0.6} $, but the relation
flattens for lower luminosity systems which have
velocity dispersions below $300~{\rm km~s^{-1}}$).

We suggest that the $L_x-\sigma$ relation arises from a combination of extended
cluster emission and emission associated with individual galaxies.
The x-ray emission for the richer groups is dominated by emission from the
intragroup medium, as for the richer clusters; emission from the poorer
clusters
 is dominated by less extended emission associated with the individual group
galaxies.
\vskip 0.6 in
{\bf 1.  INTRODUCTION}
\vskip 0.2 in
Groups of galaxies are one of the most common environments in the universe:
most galaxies are members of groups or clusters (Soneira \& Peebles, 1978,
Ramella {\it et al.} 1989, hereafter RGH).  Groups trace large-scale structure,
 and are the primary
constituents of large-scale features like the Great Wall (Geller and Huchra
1989, Ramella et al 1990).  The distribution of group velocity dispersions is
 a constraint on models for
the formation of large-scale structure
 (eg. Ueda {\it et al.} 1993, Moore {\it et al.} 1992).

Catalogs of individual groups provide information about the mass-to-light
ratios
 and evolution of systems of galaxies.  Optical data alone can be used to test
whether groups of galaxies are relaxed systems.  For example, the work of
Diaferio {\it et al.} (1993) suggests that they are
dynamically young.  X-ray data can provide a mass-to-light estimate for
groups independent of the optical data (Kriss {\it et al.} 1983, hereafter KCC,
Mulchaey {\it et al.} 1993), and
the relation between the x-ray and optical properties
 can provide further clues to the history of the groups and their present
state.

  Although there have been several studies of the optical properties of groups
(RGH, Hickson {\it et al.} 1982, Bahcall 1980, Beers {\it et al.} 1984, Huchra
\& Geller 1982, Diaferio et al 1993), and some analyses of the
x-ray emission from groups (KCC, Price {\it et al.} 1991, hereafter PBDN,
Mulchaey {\it et al.} 1993, Ponman \& Bertram 1993), there has been no attempt
to obtain complete optical data for a large set of systems observed in  x-rays.

Here we study a sample of 31 groups and poor clusters observed with the
Einstein Observatory.  We have collected 668 redshifts for galaxies in the
 group fields.
We use the redshift, galaxy surface number density, and x-ray data to
investigate the
physical properties of groups.  Throughout the paper, we assume
$H_o=100{\rm~ km~s^{-1}~Mpc^{-1}}$.

Section {\it 2.1} contains a discussion of the observations.  In section
{\it 2.2}, we discuss the group selection criteria.  Section {\it 2.3} contains
 the results of the analysis of the optical data.  Sections {\it 3.1-3.5} deal
 with the x-ray data reduction.  Finally, in section {\it 4}, we compare the
optical and x-ray data.
\vskip 0.3 in
{\bf 2.  OPTICAL DATA}
\vskip 0.2 in
{\it 2.1  Group Selection and Spectroscopy}
\vskip 0.2 in

Our sample includes 31 groups of galaxies observed with the Einstein satellite.
 Of these, 26 ``groups" are in 23 fields observed by PBDN
, and 5 were observed by KCC.
The sample is not
 statistically complete; the groups were originally
selected because they had been observed with the VLA (Burns {\it et al.} 1987),
 or because they contain a
centrally dominant galaxy (MKW-AWM groups).  A few other groups have been
observed with the Einstein satellite.  In some cases, we have omitted them
from our analysis because we do not have complete redshift information.  We
 omit
other groups because they are too nearby to fit within the Einstein field.

For each of these groups, we obtained redshifts for galaxies with $m_B\le 15.7$
 within
$1.5^{\circ}$ of the x-ray pointing centers, corresponding to a radius of $\sim
2.8$ Mpc at the redshift of the farthest group ($z\sim 0.04$), and $\sim 0.4$
 Mpc for the nearest
 one ($z\sim 0.005$).  In all but the two nearest cases, this radius is larger
than the
typical size of groups objectively selected from a redshift survey ($r_H \simeq
 0.6$ Mpc, cf. RGH).
 The angular scale is also slightly
larger than the field of view of the Einstein satellite.  We chose this angular
scale rather than a physical scale because the groups were originally
selected on the basis of their angular extent.
We examine the group properties within the $1.5^{\circ}$ field and within a
 fixed 0.7 Mpc circle around the x-ray centers.  Table 1 lists the new
heliocentric radial velocities ($cz$).

We measured redshifts of the group galaxies with the
photon-counting Reticon systems (Latham 1982) on the Tillinghast
Reflector (1.5 m) at the Whipple Observatory or with the blue channel of the
MMT spectrograph.
We obtain heliocentric radial
velocities by
cross-correlating object spectra against stellar and galaxy templates
(Tonry \& Davis 1979, Kurtz {\it et al.} 1992), or by fitting a gaussian
function to
emission lines.
The cross-correlation errors are estimated from the width of the correlation
peak
and its height relative to the noise.  If emission lines are present, we
estimate the error from
the dispersion of velocities determined from individual emission
lines, weighted by the goodness of fit (Tonry \& Davis 1979, Kurtz {\it et al.}
 1992).
The average calculated external error for our velocity measurements
is $\sim 35~{\rm km~s^{-1}}$; the actual external error should be only slightly
larger (c.f. Lewis 1983).
In all cases we quote heliocentric velocities in the form
$v=cz$, where $z$ is the measured spectral redshift.

  For each group, table 2 lists the position, mean radial velocity, and number
of velocities for galaxies brighter than $m_b=15.7$ in both the $1.5^{\circ}$
and $0.7$ Mpc samples.
 We derive these quantities from a total of 543 redshifts from the literature
 (Huchra {\it et al.} 1992)
along with our 125 new measurements.  Four of the groups, N56-388, N56-391,
N56-394a, and MKW3, are not complete to $m_B=15.5$; we omit these from
the surface number density analysis, but we include them in our x-ray
luminosity
 and
velocity dispersion studies.
  In several cases, there are
galaxies fainter than $m_B=15.7$ with known redshifts in our fields.  We
include these galaxies in our calculation of the velocity dispersion.

Column 1 of table 2 lists the name of the group; columns 2-4 contain the
right ascension of the group centers; columns 5 and 6 list
the group center declination.  Column 7 tabulates the mean recession velocity
for each group
and the uncertainty in the measurements derived from the individual measurement
 uncertainties and a statistical jack-knife procedure (Diaconis and  Efron
1983)
.  Columns 8 and 9 and columns 10 and 11 list the number of group members in
the
 1.5$^{\circ}$  and 0.7 Mpc
samples, respectively.  In each case, the first number represents the
number of group galaxies known, the second is the number of galaxies with
measured redshift and with
$m_B\le 15.7$.  Column 12 gives the completeness limit for each group.

  Figures 1a-1f show the spatial distribution of
the galaxies on the sky, and the position of the 0.7 Mpc circle for all 31
groups.  In Figures
1a-1f, the filled circles represent group members; the crosses and triangles
represent foreground and background galaxies, respectively.
\vskip 0.3 in
{\it 2.2  Group Membership}
\vskip 0.2 in
Once we have selected the group fields, we define the group ``membership".
Objective group selection algorithms extract groups from large-scale surveys by
 making cuts in projected separation and velocity (RGH).  However, one cannot
``objectively" select the groups in this way a posteriori.

  Our group fields were selected by other workers and do not represent a
complete sample.  In effect, a selection based on angular size has already been
 applied.  We thus limit ourselves wherever
possible to using the velocity separations as an additional
membership criterion.

 All groups except MKW2, N67-336a and N67-336b can be defined by a velocity cut
 alone, a consequence of the effective spatial cut made previously to
select the group fields.
We calculate
 the velocity dispersion of the central envelope of galaxies (defined as all
 galaxies differing by less than 200 ${\rm km~s^{-1}}$ from another member).
  If the
separation
between an outlier and the nearest assigned member is less than $1.2\sigma$,
 we add the galaxy to the group list.  This procedure is insensitive to
the minimum $\sigma$ criterion.  We repeat the procedure for all the
outliers.  Although our procedure differs from the $3\sigma$ clipping
procedure (Yahil \& Vidal 1977), in practice the groups are well separated
in redshift
space, and the difference between our procedure and 3$\sigma$ clipping is
generally negligible.  In only one case, S49-147, the velocity dispersion
derived by our method differs from the 3$\sigma$ clipping velocity dispersion
by
 more than the error in the estimate:  for S49-147, inclusion of the outliers
as mandated by 3$\sigma$ clipping yields a velocity dispersion $\sigma=422$
${\rm km~s^{-1}}$.  This higher velocity dispersion is caused by
symmetrically placed outliers in redshift space (see figure 2a).  These
outliers are significantly separated from the well-defined central peak;
furthermore, all the outliers lie far from the core of the group (see figure
1a).
The lower $\sigma=246 ~{\rm km~s^{-1}}$ probably makes more sense.

Even in cases where there are two groups superimposed on the
sky, our procedure is robust provided that
 the redshift separation between the groups is large enough
(cf. N56-394a and N56-394b.)
For the two cases where there are overlapping groups -- MKW2 and MKW2s, and
N67-336a and N67-336b -- we use the galaxy positions
 to make the membership assignments (we do not include MKW2S in our group
sample
 because we lack complete redshift data for it).
Because it is virtually
impossible to assign the galaxies in the overlap region, we
consider only the cores of the groups, defined as circles of $0.45$ Mpc about
the optical centers of the subclumps (corresponding to the density peaks and
listed in table 3).  Although
this procedure greatly reduces the number of redshifts available for analysis,
 it reduces the risk
of contamination from the neighboring group.

 Figures 2a-2f
show the radial velocity distributions for galaxies in our $1.5^{\circ}$
fields;  the
 group members appear as hatched histograms.
Table 4 lists the derived velocity dispersions (corrected for ``redshift
inflation" in accordance with Danese {\it et al.} 1980)  of our systems for
 both the
$1.5^{\circ}$ (columns 2 and 3) and 0.7 Mpc samples (column 4 and 5).  In each
 case, we calculate the velocity dispersion first using all the available
redshifts and then using only those of galaxies brighter than $m_B=15.7$.
Column 6 lists the group surface number density parameter $\mu$ (see section
2.3).

  With the exception of N45-389, the derived velocity dispersions
are generally quite stable, despite the large variations in the number of
galaxies in the different samples for a particular group.  In the case of
N45-389, the difference in the velocity dispersion between the samples might be
 attributable to an outer envelope of galaxies infalling onto a tightly bound,
low velocity dispersion, central core.  In other cases where the velocity
dispersion varies significantly -- N56-369, N56-394b and MKW6a -- the number of
 galaxies used to determine the dispersion is small (3 or 4 galaxies). In these
 cases the
velocity estimates in table 3 are underestimates by as much as a factor of 2,
and the 1-D velocity dispersion we compute is a biased estimator of the 3-D
velocity dispersion (Diaferio {\it et al.} 1993).
Fortunately, this bias in the computed velocity dispersions for groups with
few redshift measurements affects only a few of the groups, and thus does not
affect the conclusions of this study.
\vskip 0.3 in
{\it 2.3   Correlation of Velocity Dispersion and Surface Density}
\vskip 0.2 in
Because our groups were selected according to their angular size, the physical
scale of the groups is a function of distance.  The procedure for group
selection
 makes it difficult to define a group ``radius" a posteriori.  We do not yet
have sufficient photometric data to determine luminosity functions for the
groups.
Because of the difficulty of uniformly determining a spatial scale and the
 uncertainty
 in the optical luminosity of the groups, derived mass-to-light ratios are
indeterminate.
  We therefore focus on the velocity dispersion and surface density which we
can
 derive from our data.

Although the value for $N_{gal}$ in table 2 is related to the ``richness" of
the
groups, a proper estimate of the surface number density of bright galaxies
in these
groups requires a correction for the range in the group redshifts.
We need to normalize all the groups to same distance.
We count
group members brighter than $m_B=15.7$ within the 0.7 Mpc circle (smaller than
the 1 Mpc circle chosen by Zabludoff {\it et al.} (1993)).   We normalize to
a fixed distance by assuming that all the groups have the same Schechter
luminosity function.
We choose a Schechter luminosity
function with
$\alpha=-1.2$ and $M_*=-19.15$, and normalize our distances to 130 Mpc.
This choice agrees with the normalization of Zabludoff {\it et al.} (1993).
Where we are not complete
to $m_B=15.7$, we extrapolate from the number of galaxies brighter than
$m_B=15.5$.  To compare our data with Zabludoff {\it et al.} (1993), we
normalize by the mean ratio of the number of group members within 1.0 Mpc
and 0.7 Mpc, $n(1{\rm Mpc})/n(0.7{\rm Mpc})=r$; we use groups where we have a
complete survey to 1 Mpc and $m_B=15.7$. This ratio, $r$, is 1.42.
The number densities are not corrected to account for Galactic obscuration.

Figure 3 shows the relationship between the normalized surface density and
$\sigma$ for our groups and for the clusters of Zabludoff {\it et al.} (1993).
  The velocity dispersion is derived using all group galaxies, including those
with $m_B > 15.7$ (column 3, table 4).
The correlation between the surface density of galaxies and velocity dispersion
extends across the entire observed range of velocity dispersions
($50~ {\rm km~s^{-1}} \le \sigma \le 1200~ {\rm km~s^{-1}}$).  The best fit
relation is
$$log(\mu) = -4.3(\pm 1.6) + 1.6(\pm 0.6)log(\sigma), \eqno (1)$$
with a Q value (Press {\it et al} 1993) of 0.25 (Here Q is the probability that
 the chi-square value of the fit would occur by chance).  The fit is quite a
bit
more robust when we exclude the poor cluster AWM7.  AWM7 is the group most
affected by galactic absorption, and therefore we probably underestimate the
surface density of AWM7 relative to the other groups.  Correcting for this
effect would improve the fit.

Simple models of groups can account for the observed correlation.
For an isothermal sphere, the
integrated surface mass density inside a fixed radius follows the relation
$\mu \propto \sigma^2$
 For a King model, the relation is slightly flatter because the velocity
dispersion
 drops outside the
core of the system.  A reasonable approximation to a King model gives $\mu
\propto \sigma^{1.8-1.9}$.   For our sample of poor clusters, the observed
slope ($1.6\pm 0.6$) is
 certainly consistent with these
estimates from simple models.

\vskip 0.3 in
{\bf 3.  THE X-RAY DATA}
\vskip 0.2 in
{\it 3.1 Observations}
\vskip 0.2 in
The 31 ``groups" in our sample were observed with the Einstein Observatory
Imaging
Proportional Counter (IPC).  A total of 32 fields were observed, with
observation times varying from 400 seconds
to 23000 seconds.  For a few of the shorter observations, we derive only upper
limits for the luminosity, but many of the groups are detected (PBDN, KCC).

We calculate the x-ray luminosities from the Einstein images using the counts
in the 0.3-3.5 keV range, which corresponds to Pulse Height Invariant (PI) bins
 3-10.  We use the optical positions of the
group galaxies to check that the x-ray emission is actually associated
with the group.  To provide the best possible signal-to-noise ratio, we compute
 the luminosities only in the region where there is a detectable excess over
background determined from the radial x-ray surface brightness profile.
We thus obtain isophotal luminosities for the groups, rather than
fixed-aperture
 luminosities.
This procedure might underestimate
the total x-ray luminosity of some of the groups, particularly those with short
 observation times.

Table 4 lists the x-ray
parameters for the observed groups.   The second column lists the galactic
HI column density (in ${\rm cm^{-2}}$) in the direction of the groups, derived
from the Burstein \& Heiles (1983) maps.  Column 3 gives the area (in square
arcminutes)
 used in the flux determination.  Column 4 gives the
mean group redshift.  Column 5 gives the background-subtracted counts
measured for each group.  Column 6 lists the effective exposure time for the
Einstein observation.  The derived temperature for each group is in column 7.
Finally, Columns 8 and 9 tabulate the x-ray luminosities for each group
in units of $10^{42}h^{-1}~{\rm ergs~s^{-1}}$ for two assumed group
temperatures, 1 keV and the temperature derived from the $L_x-T$ relation (Edge
\& Stewart 1991), respectively.

     In general, the groups have x-ray luminosities
$\simless 10^{42}h^{-1}  ~{\rm ergs~s^{-1}}$.  Thus, with
very few exceptions, the extraction of the group x-ray emission from the data
is
 quite complex.  In many cases, we expect the x-ray maps to be dominated by
emission from the
individual galaxies.
 At the distances typical of the poor
clusters in the sample (60-90 Mpc), almost all the galaxies have (optical)
angular sizes less than an arcminute.  The typical resolution for the Einstein
IPC is $\sim 1.5 '$.
 Although the largest galaxies in the nearer groups will be slightly extended
when observed with the Einstein IPC, most of the galaxies under consideration
are considerably smaller than the IPC resolution; as a first approximation,
we model them as point sources.
\vskip 0.3 in
{\it 3.2  Creating a Point Response Function}
\vskip 0.2 in

Many groups show x-ray emission strongly clumped at locations corresponding
to the positions of bright galaxies.  A natural interpretation for this
phenomenon is that the x-ray emission is just the integrated emission from
individual sources
inside the galaxies themselves, or possibly from hot gas associated with the
interstellar medium of the galaxies (Fabbiano {\it et al.} 1992).
As a first step in identifying the contribution of individual galaxies to the
X-ray emission from the poor clusters, we model the emission from individual
galaxies.

The Einstein IPC has a complicated point response function, which depends on
 both the photon energy and the instrument gain.
To simulate these point sources, we follow
Mauche (1983):  we model the IPC as a
convolution of an energy-dependent mirror response function and a (gaussian)
voltage gain setting-dependent detector response.  For each typical value
of the IPC gain setting, we then can create a point
response function dependent on the energy spectrum of the incoming x-rays.
In order to estimate the effects of the detector response, we use the raw
pulse-height channel (PH) data rather than the PI binning.

Because early and late type galaxies have substantially different x-ray
spectra,
we construct two representative ``point sources" per gain setting--
one for a spiral galaxy, the other for an elliptical.
We use representative x-ray spectra from Fabbiano {\it et al.} (1992).
Given the energy band and the IPC gain
setting, we convolve a gaussian of specific width (Harnden {\it et al.} 1984)
with an Einstein HRI monochromatic test exposure.
After convolving the maps for each channel, we construct a weighted sum of the
resulting maps to obtain a model point response function.
We also construct model point sources for AGN with known spectra (Wilkes \&
Elvis 1987) in order
to test the validity of the procedure.  The resulting image fits the spatial
distribution of point sources quite well (figure 4).

The variations in the point spread functions for different spectra
are not very large.  Typically, the width of the profile  varies by $\simless
5 \%$ FWHM.
 Variations in gain are a more serious problem.
  The high voltage gain setting
influences the point spread function by changing the correspondence between
photon energies and pulse height channel.  If we consider the same energy range
 for all the maps, then we need to consider different pulse-height channels.
This problem is especially serious at the low energy end of the x-ray spectrum,
 because the difference in FWHM between the channel 2 and channel 3 IPC
response
is  $\sim 25 \%$.  Figure 5 shows the difference in the radial profiles for
identical point sources observed with gains of 12, 14, 16, 18.
We use these radial profiles to determine whether there is extended x-ray
emission associated with the galaxies in the groups.
\vskip 0.3 in
{\it 3.3  X-ray Image Reduction}
\vskip 0.2 in
	To reduce the full x-ray images, we first subtract
a standard background file, scaled according to the count rate in the image
away
 from strong sources.  We then multiply by a flat field map to correct for the
variation in detector response across the image.   We construct an error image
from the data assuming Poisson noise and an uncertainty of 25\% in the
background level determination.  Finally, to aid in the identification of low
surface brightness features, we smooth the image (and the error map) with a
1.5$^\prime$ gaussian.  We use the unsmoothed images in
all the analyses.

Because the Einstein IPC fields often contain sources not associated with the
groups, we compare the x-ray map with the observed galaxy distribution to
identify the emission associated with the group.  We then determine the number
of counts associated with the group by summing the emission from the selected
regions of the unsmoothed background-subtracted maps.
\vskip 0.3 in
{\it 3.4  Group Profiles, Poor vs. Rich Systems}
\vskip 0.2 in
It is interesting to contrast the morphological appearance of high and low
density and velocity dispersion groups.  For example,
figures 6 and 7 compare the profiles for a high velocity dispersion
(N67-335==MKW4) and a low velocity dispersion group (MKW10).
In both cases, there is a single
bright source associated with the group.  In MKW10
($\sigma=165~ {\rm km~s^{-1}}$), however, the
radial profile of the source is only slightly extended;
the excess over the point source profile (the dotted line in figure 6) accounts
for only
$\sim$12 \% of the total emission.  In contrast, N67-335
($\sigma=476~ {\rm km~s^{-1}}$) is clearly extended.
  Here the excess over the point source profile is more than a factor of
five.  The surface number density of N67-335 ($\mu =0.748$) is more than twice
that of MKW10 ($\mu = 0.305$), as expected from the $\mu - \sigma$ relation.

There is an easily discernible trend in the qualitative nature of x-ray surface
brightness distributions for groups:  groups with a higher velocity
dispersion (and a higher surface density of galaxies) have more extended
emission,
 often centered on the dominant optical galaxy.  These systems also tend to
have smoother, more regular
x-ray surface-brightness contours (AWM4, figure 8), than the less
luminous, lower velocity dispersion systems (e.g. S34-111 and S49-138, figures
9, 10).  For the lowest velocity dispersion
systems ($\sigma \simless 200~ {\rm  km~s^{-1}}$), the emission appears to be
concentrated almost
exclusively around a few bright ($m_B\le 15.7$) galaxies; the x-ray morphology
of these groups is in accordance with our model for the nature of the group
emission (see section 3.5).

The x-ray emission for low $\sigma$ groups is
 roughly consistent with that expected from the individual galaxies.
As a check on this model, we added up the best (highest signal-to-noise)
observations of groups made with high voltage gain settings of 16 and 18
(N79-286, N67-336a, N56-381, S49-138, and MKW1s).
We stack the images to superimpose the center of emission, or, in the case
where
 there seem to be distinct sources (e.g. S49-138), the centers of emission.
  Figure 11 plots the radial profile of this ``summed image" compared to the
point source profile for the same gain.  The profile is only
slightly extended, showing an excess of $\sim 10\%$ over the point source.
Some
 of this emission can be attributed to the presence of other point sources in
the vicinity.  In addition, uncertainties is centroiding the superimposed
images
 could account for some of the excess.  We estimate
that positional uncertainties contribute an uncertainty of $\sim 3\%$ in the
point source fit.

\vskip 0.3 in
{\it 3.5  X-Ray Counts-to-Flux Conversion and Temperatures}
\vskip 0.2 in
Because the energy bins are
very broad, the  x-ray counts-to-flux conversion depends on the input energy
spectrum,
and thus on the temperature of the groups.  Temperatures for the poor clusters
have not been well measured, but
ROSAT observations (Ramella 1993) indicate that the temperatures are
 $\sim 1$ KeV.

  We calculate the counts-to-flux conversion in two
ways.  First,
 we assume all the groups have a temperature of 1 KeV.
In a second approach, we extrapolate the $L_x-T$ relation for rich clusters
(Edge \& Stewart, 1991) to lower luminosities
and use the relation to derive a temperature and a luminosity for each group
iteratively.
This second procedure leads to luminosities which are up to 40\% lower at the
low luminosity end than the first approach.
Some of the poor clusters at the high velocity dispersion end of our sample do
have published X-ray temperatures (KCC, Schwartz {\it et al.} 1980).
To treat all our groups consistently, we use the temperature estimates from the
Edge \& Stewart relation instead of the measured temperatures.
These two approaches agree in all cases to within the published
error bars (KCC).

For completeness, Table 5 includes fluxes determined from both approaches;
columns 8 and 9 give the x-ray luminosities in the 0.3-3.5 KeV band:  column 8
tabulates the luminosity assuming group temperatures of 1 KeV, and column 9
contains the luminosities assuming the derived temperatures (listed in column
7)
 .
  The difference between these two values is an indication of the
systematic uncertainties in the x-ray luminosity measurements.

In figure 12, we plot the x-ray group luminosities (assuming T=1 KeV) against
the velocity dispersions; for comparison, we also include luminosities and
velocity dispersions for some rich clusters (Zabludoff {\it et al.} 1993,
Struble \& Rood 1987).  X-ray luminosity and velocity dispersion are obviously
correlated.  We find
$$log(L_x)\sim 31.81(\pm 1.67) +4.0(\pm 0.6)log(\sigma) \eqno (2)$$
 using our groups alone.  This result is consistent
with the relation derived for rich clusters (Quintana {\it et al.} 1982).
If we assume the temperatures derived from the $L_x-T$ relation for our groups,
we find
$L_x(T_g) \propto \sigma^{4.2\pm 0.7}$, not significantly different from (2).
  However, if
we only consider the groups which show emission associated with individual
sources (those with $L_x \le 1.5\times 10^{42}h^{-1}~ {\rm  erg~s^{-1}}$), we
find a shallower slope, $L_x \propto \sigma^{2.7\pm 1.3}$, albeit with
greater scatter.  This flattening occurs regardless of the $L_x-T$ relation
assumed.

Because we only calculate luminosities within regions where the emission is
stronger than the background, the
luminosities we calculate are lower limits:  typically we would not detect a
diffuse component with $L_x \simless 1-4\times 10^{40}~{\rm ergs~s^{-1}}$.
However, any additional emission only accentuates the deviation of the poor
groups from the $L_x-\sigma$ relation defined by the richer systems.

\vskip 0.3 in
{\bf 4. DISCUSSION}
\vskip 0.2 in
{\it 4.1  $L_x-\sigma$: Observation  and Theoretical Considerations}
\vskip 0.3 in
The standard explanation for the observed relation between the x-ray luminosity
and the velocity dispersion of rich systems of galaxies is that both
quantities depend on the mass of the cluster.
Thermal emission from the
intracluster gas yields an x-ray luminosity proportional to the square of the
gas density.
For a constant mass-to-light ratio, the x-ray luminosity is then
proportional to the square of the mass of the cluster.  If the cluster is a
relaxed system,
the velocity dispersion is roughly proportional to the square root of the mass.
Thus,
 $L_x \propto \sigma^4$,
consistent with the observed slope derived from our group sample
($50~ {\rm km~s^{-1}} \simless \sigma \simless 800{\rm~km~s^{-1}}$)  (figure
12)
, and the slope previously derived for clusters.

Figure 12 shows that although the rich (high dispersion) groups follow the
$L_x-\sigma$ relation defined by clusters quite closely, there appears to be a
flattening of the
relation for velocity dispersions $\simless 300~ {\rm km~s^{-1}}$.
A simple model shows that the increasing relative contribution of the
integrated
 emission from individual galaxies {\it should} cause a shallower slope for
these lower $\sigma$ systems.

For both spiral and elliptical galaxies there is a correlation between the
x-ray
 and optical luminosity (Fabbiano {\it et al.} 1992):

$$ L_x(elliptical)= 3.16\times 10^{21} L_B^{1.8} \eqno (3) $$
and
$$ L_x(spiral)=2.00\times 10^{29}  L_B^{1.0} \eqno (4) $$
where $L_x$ is in ${\rm erg~s^{-1}}$ and $L_B$ is in solar luminosities.
There is a large scatter
(almost an order of magnitude!) around these relations.
If the group luminosity function is a Schechter luminosity function, the
total x-ray luminosity from the group is
$$ L_x(tot)=r\mu_o \int_{Xmin}^{\infty} x^{-\alpha}e^{-x}(f_{sp}L_x(spi,x)+
(1-f_{sp})L_x(ell,x)) \eqno (5) $$
where $x =L/L_*$, $r$ is the correction factor from section 2.3, and $\mu_o$ is
 a fiducial surface number density.  The lower limit of integration is
$X_{min}=L_{min}/L_*$ and $L_{min}$ is the minimum galaxy luminosity (which we
take to
be $M_B=-13$.)

 Our measured
surface number density is
$$ \mu = \mu_o\int_{Xo}^{\infty} x^{-\alpha}e^{-x}dx  \eqno (6) $$
where $X_o=L_o/L_*$ and $L_o$ is the luminosity corresponding to our limiting
magnitude of $m_B=15.7$ at a distance of $130$ Mpc ($M_o = -20.19$).
We can then write (3) in terms of $\mu$:
$$ L_x(tot)= r\mu \left( {\int x^{-\alpha}e^{-x}(f_{sp}L_x(spi,x)+(1-f_{sp})
L_x(ell,x))\over \int x^{-\alpha}e^{-x}dx }\right). \eqno (7) $$
This expression can be integrated numerically given $\alpha=-1.2$ and
$L_*=3.5\times 10^9L_{\odot}$
(see section 2.3).
To solve for $L_x$ we need a value for the spiral fraction  $ f_{sp}$:  we use
a rough median value for the spiral fraction observed in our groups
$f_{spi}=0.66$.  The range of spiral fractions in
our groups introduces an extra uncertainty in the $L_x-\mu$ relation.
Calculations show that varying $f_{sp}$ from 0.4 to 0.9 introduces a factor of
3 variation in the normalization of the $L_x-\mu$ relation.

In order to relate this theoretical prediction to a model of the $L_x-\sigma$
relation, we first fix the normalization of the $\mu-\sigma$ relation using
our
observed correlation (equation (1)).
  Thus for groups where the galaxy contribution is dominant, we expect
$L_x \sim 4.5\times 10^{36} \sigma^{1.6} {\rm ~ergs~s^{-1}}$.
The dotted lines in figure 11 show the range of luminosities expected as a
function of $\sigma$, taking
into account the observed scatter in the $L_x -L_{opt}$ relation for galaxies,
and the uncertainty in spiral fraction.
The flattening of the $L_x-\sigma$ relation for our groups occurs just where
emission from
individual galaxies should become important---at
$\sigma \sim 300~ {\rm  km~s^{-1}}$.

We construct our heuristic model by comparing the group data with that for rich
clusters at high $\sigma$, and with x-ray properties of individual galaxies at
low $\sigma$.  In this picture,
groups of galaxies should show intermediate properties, with a transition
between ``galaxy-dominated" and ``intra-cluster medium dominated" properties
occurring
at $\sigma \sim 300 ~{\rm km~s^{-1}}$.  It is still a challenge to understand
what (if any) underlying physics
determines this transition.

\vfill
\eject
{\bf 5.  CONCLUSIONS}
\vskip 0.2 in
We have collected a uniform data set, consisting of velocity dispersions,
surface number densities, and x-ray luminosities for 31 groups of galaxies.
We find a relation between the surface galaxy number density and velocity
dispersion, $\mu \propto \sigma^{1.6\pm 0.6}$.  For comparison, for an
isothermal sphere the relation is $\mu \propto \sigma^2$, and for a King model
 $\mu \propto \sigma^{1.8-1.9}$.

The x-ray luminosity - velocity dispersion relation for rich clusters
continues to poorer systems as well, with $L_x \propto \sigma^{4.0\pm 0.6}$.
For the groups with $\sigma \simless 300~ {\rm km~s^{-1}}$, however, there is
a significant flattening in the
relation: $L_x \propto \sigma^{2.7\pm 0.3}$.
We suggest that this change in slope represents the transition between
``ICM dominated" and ``galaxy" dominated systems.  This model is consistent
with the x-ray morphology of the groups; for $\sigma \ge 300~ {\rm km~s^{-1}}$
the
groups have smooth, extended x-ray surface brightness profiles.  For
$\sigma \le 300~ {\rm km~s^{-1}}$ the group emission is consistent with a
collection of
``point sources" associated with the individual galaxies in the groups.
Unfortunately, the x-ray data available for the poor groups is not sufficient
to determine whether the intracluster medium of these groups is less abundant,
or is just too cool and/or diffuse to be seen.  Longer observations with
imaging
x-ray telescopes such as ROSAT or AXAF should allow us to answer this
question.
Because we only calculate luminosities in regions where the emission is greater
than the background, we might be underestimating the luminosity of our faintest
groups.  However, this only increases the flattening of the $L_x-\sigma$
relation, because the error does not depend on the group luminosity.

It would be interesting to investigate whether there are other quantities
which show differences between high velocity dispersion and low velocity
dispersion systems. For example,
studies of the baryon content of the Coma cluster (White {\it et al.} 1993)
suggest that the baryon fraction in clusters is substantially higher than
expected from primordial nucleosynthesis limits.  Blumenthal {\it et al.}
(1984)
 have suggested that richer MKW/AWM groups also have higher baryon fractions.
However, Mulchaey {\it et al.} (1993) have found a much lower baryon fraction
for at least one
poorer group.  It would be useful to extend
this calculation to smaller systems
by acquiring accurate photometric data to improve
estimates of mass-to-light ratios.  Detailed photometric studies along with
redshifts for even fainter members would yield information about the baryon
fraction in
groups, a potentially important constraint on models of galaxy and structure
formation
(White {\it et al.} 1993).

We thank Massimo Ramella, Antonaldo Diaferio, Joe Mohr and Ann Zabludoff for
helpful and stimulating discussions.  We thank M. Ramella for providing
information
about ROSAT observations in advance of publication.  We thank S. Tokarz for
help
 with the data reduction.  We thank the anonymous referee for useful critcism
and suggestions.
 This research funded in part by NASA grant NAGW-201, and by the Smithsonian
Institution.
\vfill
\eject
\centerline {\bf Figure Captions}
\vskip 0.5 in
 Figures 1a-f:  The positions of galaxies in the $1.5^{\circ}$ fields;
group members are denoted by filled circles, foreground galaxies by crosses,
and
 background galaxies by open triangles.  The large circles indicate regions
with
in 0.7 Mpc of the x-ray center.
\vskip 0.2 in
Figures 2a-f:  Velocity histograms for the $1.5^{\circ}$ group fields.  The
hatched
histogram represents the group members, the open histogram shows the
interlopers (or, in the case of N67-336a, N67-336b and MKW2, the galaxies not
in the core of the group).
\vskip 0.2 in
Figure 3:  A plot of the surface density of galaxies (section 2.3) versus
velocity dispersion for the groups in our sample (filled circles)
and rich clusters from the Zabludoff {\it et al.} (1993) sample (stars).
\vskip 0.2 in
Figure 4:  A comparison of the radial profile of 3C273 as observed by Einstein
with the model profile for a AGN-spectrum point source.
\vskip 0.2 in
Figure 5:  A comparison of model point-source profiles for an elliptical
galaxy measured at 4 different gain values:  18 (dashed dotted line), 16 (large
 dashed line), 14 (short dashed line) and 12 (solid line).  The radius is in
units of 8-arcsecond pixels.
\vskip 0.2 in
Figure 6:  Radial profile for the x-ray emission from the group MKW10 (squares)
and a point source model (elliptical) normalized to match at a radius of 24
arcseconds.  The error bars are 1-$\sigma$ errors.  The excess above the point
source model at larger radii accounts for $\sim 12 \%$ of the total flux.
\vskip 0.2 in
Figure 7: Radial profile for the emission from the group N67-335 (=MKW4) and
for a point source model normalized as in figure 6.
\vskip 0.2 in
Figure 8:  A contour plot of the Einstein x-ray map for the group AWM4.  The
contour levels represent  8, 15, 30 ,60, and 90 percent of the peak intensity,
 respectively.  The crosses represent the optical positions of
$m_{B(0)}\le 15.7$ galaxies in the
group field; the $x$ marks the position of the cD galaxy.
\vskip 0.2 in
Figure 9:  A contour plot of the Einstein x-ray map for the group S34-111.  The
contour levels represent 30, 60 and 90 percent of the peak intensity.
The clump of emission centered around pixel (250,100) is associated
with a background galaxy.
\vskip 0.2 in
Figure 10:  A contour plot of the Einstein x-ray map for the group S49-138.
The contour levels represent 15, 30, 60 and 90 percent of the peak intensity.
The emission appears to be concentrated around the three central
galaxies.
\vskip 0.2 in
Figure 11:  radial profile of the combined x-ray emission for
lower dispersion groups ($\sigma \simless 200~ {\rm km~s^{-1}}$) in our study.
\vskip 0.2 in
Figure 12:  A plot of the x-ray luminosity (for $T_g=1$ Kev)
versus velocity dispersion $\sigma$ for our groups (filled circles), rich
clusters
from figure 3 (filled stars) and other clusters from Rood \& Struble (1987)
(open stars).  The five-pointed stars show Rosat observations of groups
by Mulchaey {\it et al.} (1993) and Ponman \& Bertram (1993).
The dotted and solid lines outline the range of luminosities expected from the
integrated emission from individual galaxies assuming $\mu =
3.5\times 10^{38}\sigma^{1.6}$, and $\mu = 3.6\times 10^{34}\sigma^{1.6}$,
respectively.

\vfill
\eject
\vskip 0.3 in
\centerline{\bf References}
\vskip 0.3 in
\refitem Albert, C.E., White, R.A., \& Morgan, W.W., 1977, ApJ, 211, 309

\refitem Bahcall, N.A., 1980, ApJ, 328, L117

\refitem Beers, T.C., Geller, M.J., Huchra, J.P., Latham, D.W., \& Davis R.J.,
ApJ, 283, 33

\refitem Blumenthal G.R., Faber, S.M., Primack, J.R., \& Rees, M.J., 1984,
Nature, 311, 517.

\refitem Burns, J.O., Hanisch, R.J., White, R.A., Nelson, E.R., Morrisette,
K.A.
, \& Moody, J.W., 1987, AJ, 94, 587.

\refitem Burstein, D., \& Heiles, C., 1982, AJ, 87, 1165

\refitem Danese, L., De Zotti, G., \& di Tullio, G., 1980, A\&A, 82, 322

\refitem Diaconis, P., \& Efron, B., 1983, Scientific American, 248, 116

\refitem Diaferio, A., Ramella, M., \& Geller, M.J., 1993, in preparation.

\refitem Edge, A.C., \& Stewart, G.C., 1991, MNRAS, 252, 414

\refitem Fabbiano, G., Kim, D.-W., \& Trinchieri, G., 1992, ApJ Supplement, 80,
 531

\refitem Geller, M.J., \& Huchra, J.P., 1989, Science, 246, 897

\refitem Harnden, Jr., F.R., Fabricant, D.G., Harris, D.E., \& Schwarz, J.,
1984
, SAO Special Report, 393

\refitem Hickson, P., 1982, ApJ, 255, 382.

\refitem Huchra, J.P, \& Geller, M.J., 1982, ApJ, 257, 423

\refitem Huchra, J.P., Davis, M., Tonry, J., \& Latham, D., 1983, ApJ
Supplement
, 52, 89

\refitem Huchra, J., Geller, M., Clemens, C., Tokarz, S., \& Michel, A., 1992,
 Bull. CDS, 41, 31, 1992.

\refitem Kriss, G.A., Cioffi, D.F., \& Canizares, C.R., 1983, ApJ, 272, 439

\refitem Kurtz, M.J., Mink, D.J., Wyatt, W.F., Fabricant, D.G., Torres, G.,
Kriss, G.A., \& Tonry, J.L., 1992, Astr. Soc. Pac. Conf. Series, 25, 432

\refitem Latham, D., 1982, Instrumentation for Astronomy with Large Optical
Telescopes, IAU Colloquium No. 67, edited by C.M. Humphries, (Reidel
Dordrecht),
 p. 251

\refitem Lewis, B.M., 1983, AJ, 88, 1695

\refitem Malumuth, E.M. \& Kriss, G.A., 1986, ApJ, 308, 10.

\refitem Mauche, C., 1985,  Ph.D. Thesis, Harvard University.

\refitem Moore, B., Frenk, C.S., Weinberg, D.H., Saunders, W., Lawrence, A.,
Ellis, R.S., Kaiser, N., Efstathiou, G., \& Rowan-Robinson, M., 1992, MNRAS,
256
, 477

\refitem Morgan, W.W., Kayser, S., \& White, R.A., 1975, ApJ, 199, 545

\refitem Mulchaey, J.S., Davis,  D.S., Mushotzky, R.F., \& Burstein, D., 1993,
preprint

\refitem Ponman, T.J., \& Bertram, D., 1993, preprint

\refitem Press W.H., Teukolsky, S.A., Vetterling, W.T., Fannery, B.P., 1992,
{\bf Numerical Recipes in Fortran}, Cambridge University Press, 657.

\refitem Price, R., Burns, J.O., Duric, N., \& Newberry, M.V., AJ, 102, 14

\refitem Quintana, H., \& Melnick, J., 1982, AJ, 87, 972

\refitem Ramella, M., Geller, M.J., \& Huchra, J.P., 1989, ApJ, 344, 57

\refitem Ramella, M., Geller, M.J., \& Huchra, J.P., 1990, ApJ, 353, 51

\refitem Ramella, M., 1993, private communication

\refitem Soneira, R.M., \& Peebles,  P.J.E., 1978, ApJ, 211, 1
\vfill
\eject
\refitem Struble, M.F., \& Rood, H.J., 1987, ApJ Supplement, 63, 543

\refitem Schwartz, D.A., Davis, M., Doxsey, R.E., Griffiths, R.E., Huchra,
J.P.,
 Johnston, M.D., Mushotzky, R.F., Swank, J., \& Tonry, J., ApJ, 238, L53.

\refitem Tonry J, \& Davis, M., 1979, AJ, 84, 1511

\refitem Ueda, H., Itoh, M.,  \& Suto, Y., 1993, ApJ, 408,  3

\refitem White S.D.M., Navarro, J.F., Evrard, A.E., \& Frenk, C.S., 1993,
preprint

\refitem Wilkes, B.J., \& Elvis, M., 1987, ApJ, 323, 243

\refitem Yahil, A., \& Vidal, N.V., 1977, ApJ, 214, 347

\refitem Zabludoff, A.I., Geller, M.J., Huchra J.P., \& Vogeley, M.S., 1993,
ApJ
, In Press
\vskip 0.3 in
\vfil
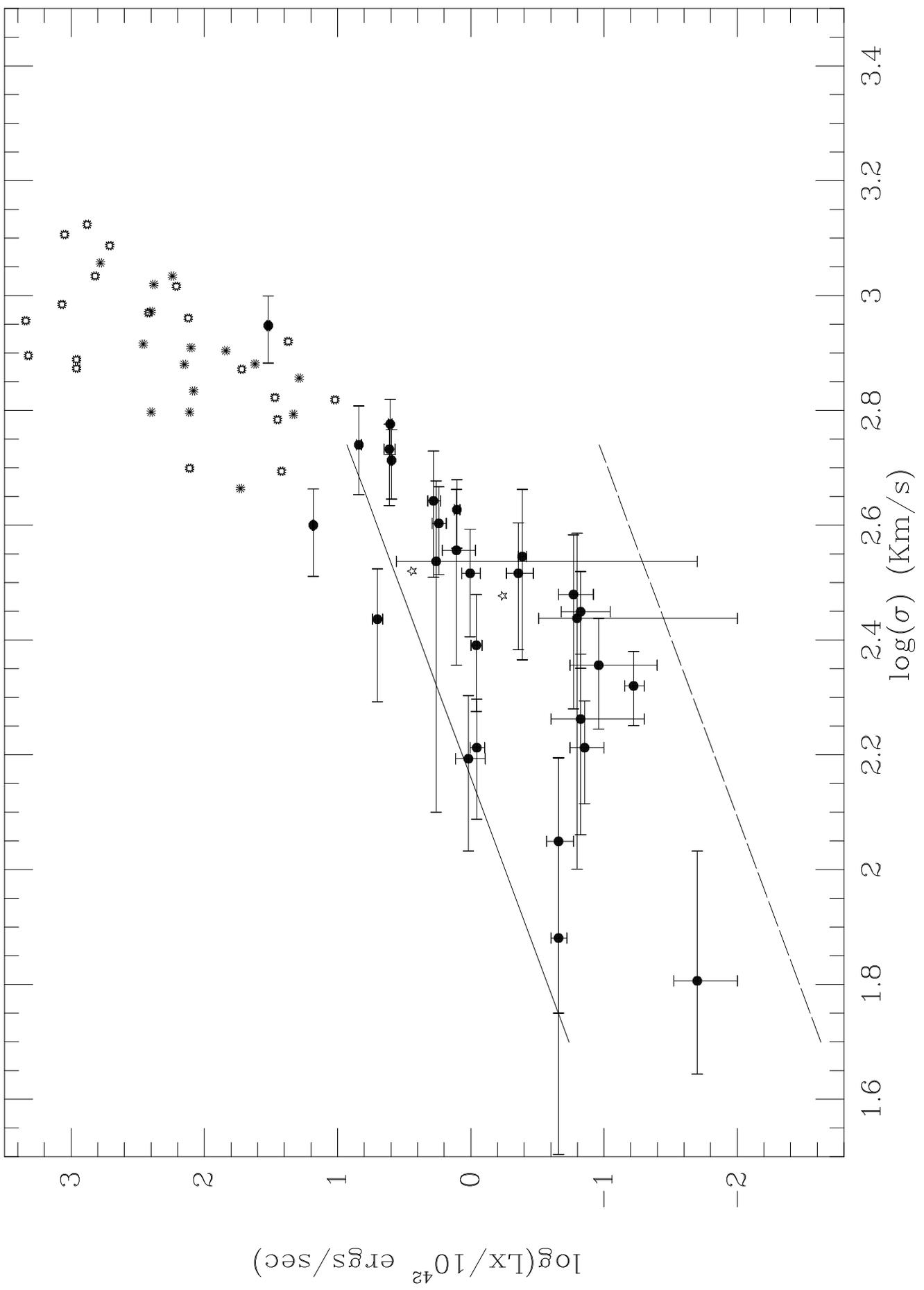
\end